\documentclass[aip]{revtex4-1}
\usepackage{graphicx}
\usepackage{amsmath}
\usepackage{xcolor}

\begin{document}


\title{Ion Beam Instabilities during Solar Flare Energy Release} 




\author{A. Fitzmaurice}
\email{afitz@umd.edu}
\affiliation{Institute for Research in Electronics and Applied Physics, University of Maryland, College Park, Maryland 20742, USA}
\author{J. F. Drake}
\email{drake@umd.edu}
\affiliation{Institute for Research in Electronics and Applied Physics, University of Maryland, College Park, Maryland 20742, USA}
\author{M. Swisdak}
\email{swisdak@umd.edu}
\affiliation{Institute for Research in Electronics and Applied Physics, University of Maryland, College Park, Maryland 20742, USA}


\date{\today}

\begin{abstract}
The linear stability of waves driven by ion beams produced during solar flare energy release are 
explored to assess their role in driving abundance enhancements in minority species such as $^3$He and 
in controlling, through pitch-angle scattering, proton/alpha confinement during energy release. The Arbitrary Linear 
Plasma Solver (ALPS) is used to solve the linear dispersion relation 
for a population of energetic, reconnection-accelerated protons streaming into a cold 
background plasma. We assume equal densities of the two populations, 
using an anisotropic ($T_\parallel/T_\perp = 10$), one-sided kappa distribution 
for the energetic streaming population and a cold Maxwellian for the background.
We find two unstable modes with parallel propagation: a right-handed wave with a frequency of the 
order of the proton cyclotron frequency 
($\Omega_{cp}$) and a left-handed, lower frequency mode. We also find highly oblique modes with frequencies below $\Omega_{cp}$ that are unstable for higher beam energies. Through resonant interactions, all three modes will contribute to 
the scattering of the high-energy protons, thereby limiting their transport out of the flare-acceleration region.
The higher-frequency oblique mode, which can be characterized as a kinetic Alfv\'en wave, will preferentially heat $^3$He, making it a
good candidate for the driver of the abundance enhancements commonly observed for this species 
in impulsive events.
\end{abstract}


\maketitle 


\section{Introduction}\label{sec:intro}

In collisionless plasmas such as the solar wind and solar corona, velocity distributions 
frequently exhibit non-Maxwellian features that can trigger the growth of plasma waves. 
Free energy is converted into heat as particles are scattered by these waves through 
resonant interactions. This has recently been observed by Parker 
Solar Probe in the solar wind, where ion-scale wave activity concurrent with proton beams 
led to scattering of the beams in velocity space perpendicular to the background magnetic field. 
\cite{verniero_parker_2020,verniero_strong_2022} 

The linear theory of ion beam instabilities using Maxwellian distributions has been 
explored extensively. Much of the early work 
\cite{montgomery_electromagnetic_1975,montgomery_electromagnetic_1976,
gary_electromagnetic_1984,gary_electromagnetic_1991,gary_theory_1993} focused primarily 
on modes propagating parallel/anti-parallel to the magnetic field, specifically the 
right-hand polarized mode that is resonant with the beam population. However, 
\citet{daughton_electromagnetic_1998} found additional left-hand polarized oblique modes 
that are dominant for beams with large densities and moderate drift speeds 
($1 \leq v_D/v_A \leq 2$, where $v_A$ is the Alfv\'en speed). \citet{voitenko_excitation_2002} and 
\citet{barik_kinetic_2019} have also shown that ion beams can generate highly oblique kinetic 
Alfv\'en waves (the form of the classic Alfv\'en wave when $k_\perp \rho_s = k_\perp c_s/\Omega_{ci}\sim 1$, 
where $k_\perp$ is the perpendicular wavenumber, $c_s$ is the sound speed, and $\Omega_{ci}$ is the ion 
cyclotron frequency \cite{hasegawa_kinetic_1977, hollweg_kinetic_1999, zhao_properties_2014}).

While much of this previous work has focused on the solar wind, ion beam instabilities 
should also be present during solar flare energy release as accelerated particles interact with 
less energetic plasma in the corona. Unlike the typical Maxwellians that have been used to identify 
instabilities in the past, ion energy spectra from both solar 
energetic particle observations \cite{reames_energy_1997,mason_3he-rich_2007} and 
reconnection simulations \cite{yin_simultaneous_2024} exhibit non-thermal power-law tails 
that can extend out to several MeV. These are better modeled by kappa 
distributions, a difference that can impact the growth rates of linear 
instabilities. \cite{shaaban_electromagnetic_2020}

Waves generated by reconnection-accelerated particles are of particular interest in the 
case of impulsive solar energetic particle events, which frequently exhibit enhancements of 
the $^3$He/$^4$He abundance ratio by up to a factor of $10^4$. 
\cite{reames_particle_1999,mason_3he-rich_2007} It is commonly believed that these 
enhancements are caused by preferential heating and acceleration through cyclotron 
resonance, due to the unique charge-to-mass ratio of fully ionized $^3$He, $q/m = 2/3$ 
(when normalized to that of protons). 
\cite{fisk_3he-rich_1978} However, the source of the waves responsible for this 
acceleration remains unknown. While many previous theories have focused on electrons as the 
driver, \cite{temerin_production_1992, roth_enrichment_1997, zhang_solar_1999, 
paesold_acceleration_2003} simulations show that the ions gain more energy during 
reconnection,\cite{eastwood_energy_2013,yin_simultaneous_2024} making them a more likely driver 
for the waves causing extreme enhancements.

Using particle-in-cell simulations, \citet{fitzmaurice_wave_2024} (hereafter denoted as FDS) explored the waves 
generated by flare-accelerated proton and alpha particles, modeled by one-sided kappa 
distributions streaming into a cold background plasma. The distributions were found to be 
unstable to parallel, right-handed waves at all beam energies explored and additional 
left-handed and oblique waves at higher energies. These waves are expected to have a significant impact on 
the particle dynamics in solar eruptions, as they were shown to both efficiently scatter 
ion beams into more isotropic distributions and increase the temperatures of $^3$He by a 
factor of 20. FDS\cite{fitzmaurice_wave_2024} proposed that waves generated by flare-accelerated proton 
and alpha beams will heat $^3$He in the regions surrounding the flare 
site. The heated particles will then stream into the flare acceleration region and increase the 
abundances there, leading to the enhancements commonly observed in impulsive events.

Due to the complexity of the initial distribution functions, FDS\cite{fitzmaurice_wave_2024} 
did not solve for the linear dispersion relation and instead relied on comparing 
simulation results to the previous linear analysis done with Maxwellian distributions. In 
this paper, we use the {\tt Arbitrary Linear Plasma Solver (ALPS)} 
\cite{verscharen_alps_2018,alps_2023_8075682} to find the unstable linear wave modes directly from the 
distributions used in FDS.\cite{fitzmaurice_wave_2024} The method for solving for the linear 
dispersion relation is explained further in Sec. \ref{sec:disp}, results from ALPS are 
presented in Sec. \ref{sec:results}, and we conclude with comparisons to the simulation 
results from FDS\cite{fitzmaurice_wave_2024} and implications for $^3$He acceleration in 
Sec. \ref{sec:disc}.

\section{Solving for the Linear Dispersion Relation}\label{sec:disp}

As discussed in \citet{stix_waves_1992}, solving for the linear wave modes of a plasma 
involves finding values of the wave frequency $\omega$ and wavevector $\Vec{k}$ for which 
the determinant of the dispersion tensor $|D|$ goes to zero. In general, $\omega$ is 
complex, with Re($\omega$) = $\omega_r$ corresponding to the real frequency of the wave and 
Im($\omega$) $= \gamma$ corresponding to growth ($\gamma > 0$) or damping ($\gamma < 0$) 
of the mode. We take the components of $\Vec{k}$ to be positive so that the direction of 
propagation is denoted by the sign of the real frequency. 

Each $\omega$ and $\Vec{k}$ solution has a corresponding eigenvector that gives the 
electric field components of the mode. In the case of electromagnetic waves, a 
polarization can be defined with respect to the background magnetic field. We take 
$\Vec{B} = B_0\Vec{z}$ so that the polarization is: 
\begin{equation*}
    P = \frac{E_y}{iE_x} \frac{\omega_r}{|\omega_r|}
\end{equation*}
In this case, $P > 0$ corresponds to right-handed modes and $P < 0$ corresponds to 
left-handed modes, with $P = \pm 1$ corresponding to circular polarization. 

Evaluating the integrals required to calculate $|D|$ can often be difficult. Therefore many 
numerical solvers approximate the distribution function as a series of Maxwellians or 
kappa functions, for which the calculations are greatly simplified. However, as shown in 
\citet{Walters_effects_2023}, these approximations can lead to significant differences in 
determining the unstable modes of a plasma.

To avoid these issues, {\tt ALPS} solves for the dispersion relation directly from 
arbitrary gyrotropic distribution functions. To begin, the distribution functions are discretized
in momentum space according to a user-defined grid. The discretized distributions are then 
used for the numerical integration, including near poles with $\gamma > 0$. In the case 
of poles with $\gamma \leq 0$, an analytic continuation is required and the distribution 
near the poles is approximated using fit functions. 

For our analysis, we use the same distribution functions as in FDS.\cite{fitzmaurice_wave_2024} 
The distribution functions for the protons and alphas (when present) consist of equal densities of a cold, 
Maxwellian population and a hot, streaming population represented by a one-sided kappa function. 
We use a 241 x 121 momentum grid with $-m_ic \leq p_\parallel \leq m_ic$ and 
$0 \leq p_\perp \leq m_ic$, where $c=20v_A$, and generate the distributions using the function, 

\begin{equation}
\label{eq:distfunc}
\begin{split}
    f & = \frac{n_c}{(2\pi m_iT_c)^{3/2}}\exp{\left[\frac{-(p_\parallel^2 + 
    p_\perp^2)}{2m_iT_c}\right]} \\ & + \frac{1}{4}(\tanh{(10p_\parallel)} + 1 )
    (1-\tanh{(10(|\Vec{p}|-0.81m_ic))})\\ 
     & \times \frac{\Gamma(\kappa+1)}{\Gamma(\kappa - 1/2)(\kappa-3/2)^{3/2}}
     \frac{n_\kappa}{\sqrt{8\pi^3m_i^3T_{\kappa,\perp}^2
    T_{\kappa,\parallel}}}\\
    & \times \left(1 + \frac{p_\parallel^2}{2m_i(\kappa-3/2)T_{\kappa,\parallel}}
    + \frac{p_\perp^2}{2m_i(\kappa-3/2)T_{\kappa,\perp}}\right)^{-(\kappa+1)}
\end{split}
\end{equation}

\begin{figure}[ht]
    \includegraphics{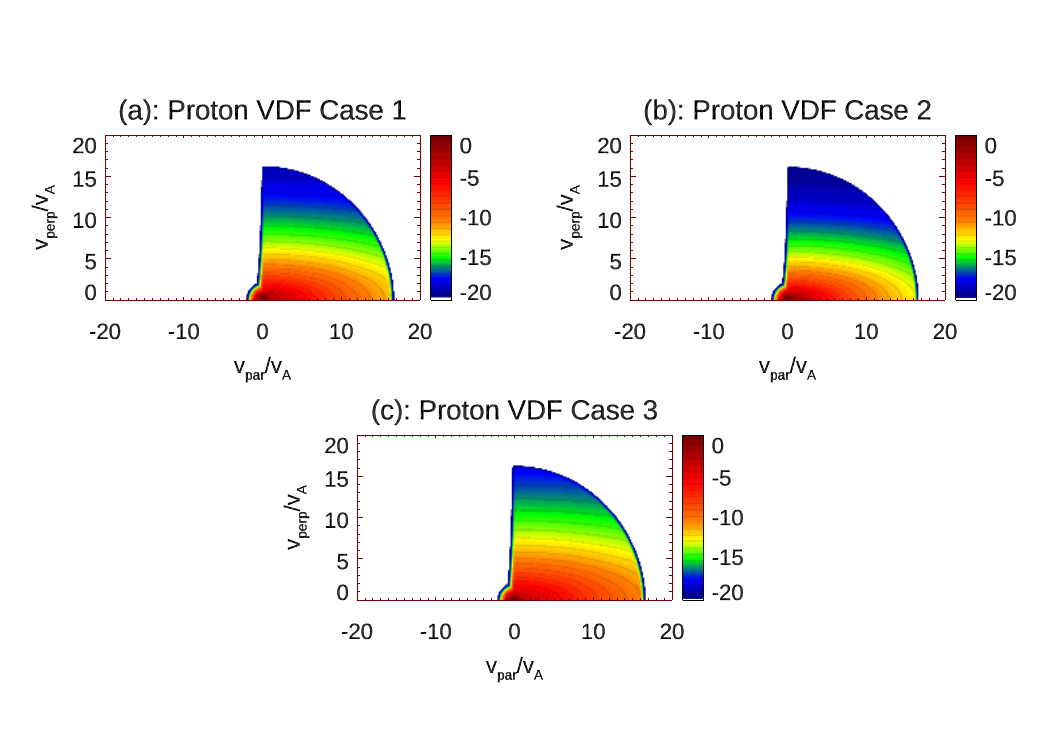}
    \caption{The velocity ($v_\parallel$ vs. $v_\perp$) distribution functions used 
    by ALPS for the protons in (a) Case 1, (b) Case 2, and (c) Case 3. The beam energy is 
    lower in Case 2 than in Case 1 and is higher in Case 3. All distributions 
    are plotted on the same logarithmic scale.}
    \label{fig:dist}
\end{figure}

The normalization parameters $n_c$ and $n_\kappa$ are set to ensure that the background and 
streaming populations each have densities equal to 0.5 and the tanh functions create cutoffs at 
$p_\parallel = 0$ and $|v|/c = 0.81$. The latter cutoff, which excludes particles near the speed of 
light, was necessary to avoid numerical inaccuracies in the PIC simulations but does not affect the 
linear results. 
As in FDS,\cite{fitzmaurice_wave_2024} we consider three initial energies for the streaming 
population. In Case 1, $T_{\kappa,\parallel} = 10 m_pv_A^2$. Case 2 has a lower initial 
energy, with $T_{\kappa,\parallel} = 5 m_pv_A^2$, and Case 3 has a higher initial energy, 
with $T_{\kappa,\parallel} = 20 m_pv_A^2$. In all cases, $\kappa = 2.5$, 
$T_{\kappa,\parallel}/T_{\kappa,\perp} = 10$ and the Maxwellian population has a 
temperature $T_c = 0.1m_pv_A^2$. Proton distributions are shown in Figure \ref{fig:dist}. In 
all cases, the distribution for the electrons is a Maxwellian with $T_e = 2m_pv_A^2$ and 
a small drift to balance the current. 

We initially consider just protons and electrons, so 
that the number densities are $n_p = n_e = 1$. We then add the alphas at 5\% the proton 
number density, so that $n_p = 1$, $n_{4He} = 0.05$, and $n_e = 1.1$. For all species, the 
distributions are treated as non-relativistic and the fit functions used in the numerical 
integration are the analytic functions used to generate the distributions.

\section{Results}\label{sec:results}
\subsection{Parallel Modes}\label{sec:parmodes}

\begin{figure}[ht]
    \includegraphics{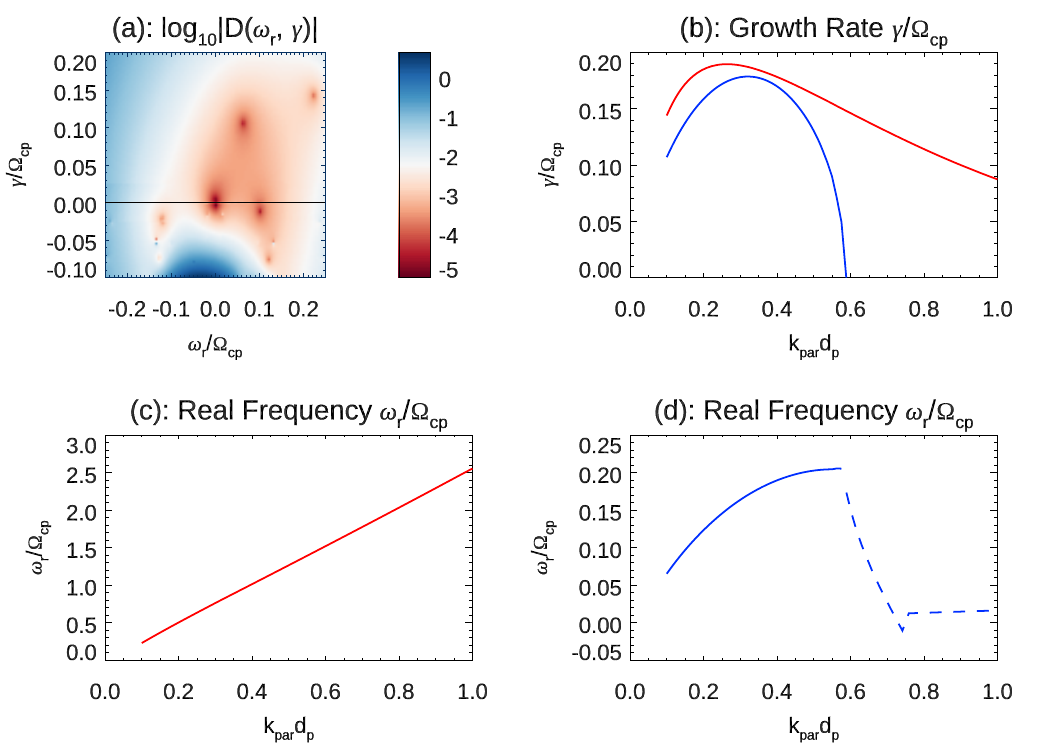}
    \caption{(a) The map of $|D(\omega_r,\gamma)|$ for Case 1 at $k_\parallel = 0.1$, 
    $k_\perp = 10^{-4}$, with minima plotted in dark red. The right-handed mode is 
    located at ($\omega_r/\Omega_{cp}$, $\gamma/\Omega_{cp}$) = (0.23, 0.14) and the left-handed 
    mode is located at ($\omega_r/\Omega_{cp}$, $\gamma/\Omega_{cp}$) = (0.065, 0.11). The (b) 
    growth rates and (c)-(d) real frequencies of the two unstable solutions are plotted as 
    functions of $k_\parallel$, with the right-handed mode plotted in red and the left-handed mode 
    plotted in blue. Real frequencies while the mode is unstable are plotted as a solid line, 
    while real frequencies while the mode is stable are plotted as a dashed line.}
    \label{fig:parmodes}
\end{figure}

We begin our analysis by looking for the unstable 
$\Vec{k} \times \Vec{B} = 0$ modes for Case 1 in an electron-proton plasma. To do this, {\tt ALPS} first 
calculates $|D(\omega_r,\gamma)|$ in the range $\omega_r/\Omega_{cp} = [-0.25,0.25]$ and 
$\gamma/\Omega_{cp} = [-0.1,0.2]$ for $k_{\parallel}d_p = 0.1$ (Fig.~\ref{fig:parmodes}(a)). 
After identifying any roots with $\gamma > 0$, {\tt ALPS} performs root-finding scans over 
$k_{\parallel}d_p = [0.1,1]$ to determine the dispersion properties of each mode (Fig.
\ref{fig:parmodes}(b)-(d)). For each solution along the wavenumber scan, the components of 
the wave electric field are also calculated to determine the polarization. 

In Case 1, we find two unstable parallel modes for $k_\parallel = 0.1$, located at 
($\omega_r/\Omega_{cp}$, $\gamma/\Omega_{cp}$) = (0.23, 0.14) and (0.065, 0.11). The 
first, higher frequency mode is a right-hand circularly polarized wave that is
non-dispersive with $\omega_r/k = 2.5v_A$. It is unstable over the full range of $k_\parallel$ 
and has a peak growth rate of $\gamma/\Omega_{cp} = 0.19$ at $k_{\parallel}d_p = 0.27$. The 
second, lower frequency mode is a left-hand circularly polarized wave that is dispersive, 
with the frequency plateauing around $0.2 \Omega_{cp}$ before becoming stable above 
$k_{\parallel}d_p = 0.6$. The mode reaches a peak growth rate of 
$\gamma/\Omega_{cp} = 0.18$ at $k_{\parallel}d_p = 0.32$, $\omega_r/\Omega_{cp} = 0.17$.

\begin{figure}[ht]
    \includegraphics{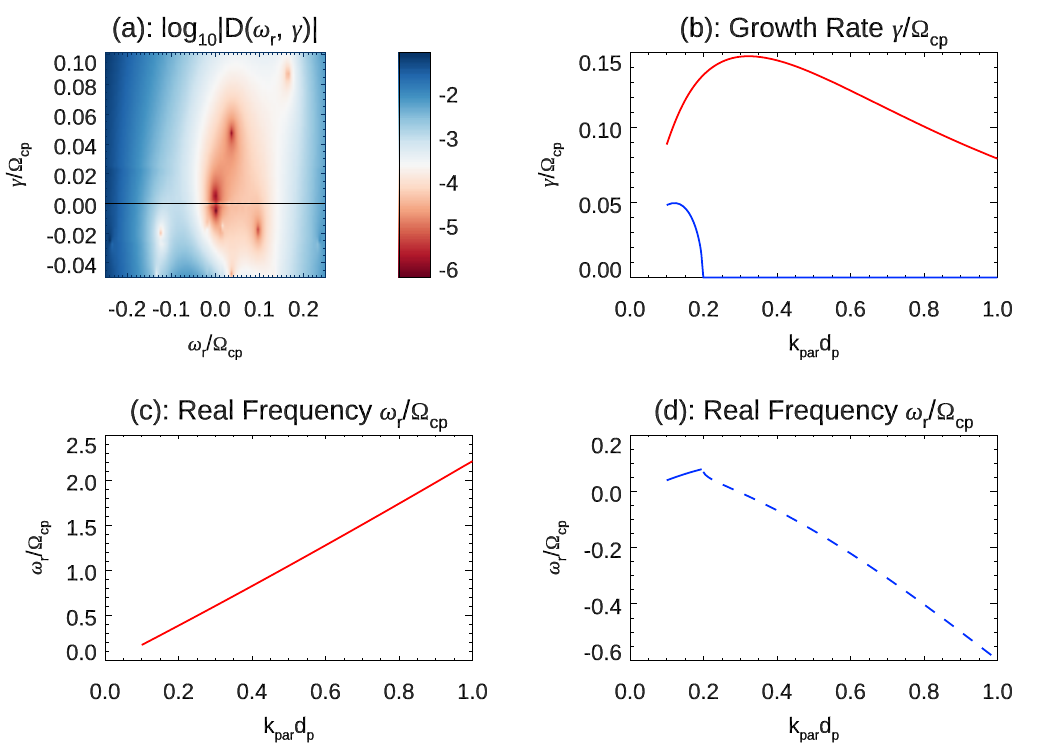}
    \caption{(a) The map of $|D(\omega_r,\gamma)|$ for Case 2 at $k_\parallel = 0.1$, 
    $k_\perp = 10^{-4}$, with minima plotted in dark red. The right-handed mode is 
    located at ($\omega_r/\Omega_{cp}$, $\gamma/\Omega_{cp}$) = (0.17, 0.089) and the left-handed 
    mode is located at ($\omega_r/\Omega_{cp}$, $\gamma/\Omega_{cp}$) = (0.039, 0.048). The (b) 
    growth rates and (c)-(d) real frequencies of the two unstable solutions are plotted as 
    functions of $k_\parallel$, with the right-handed mode plotted in red and the left-handed mode 
    plotted in blue. Real frequencies while the mode is unstable are plotted as a solid line, 
    while real frequencies while the mode is stable are plotted as a dashed line.}
    \label{fig:parmodes_cold}
\end{figure}

\begin{figure}[ht]
    \includegraphics{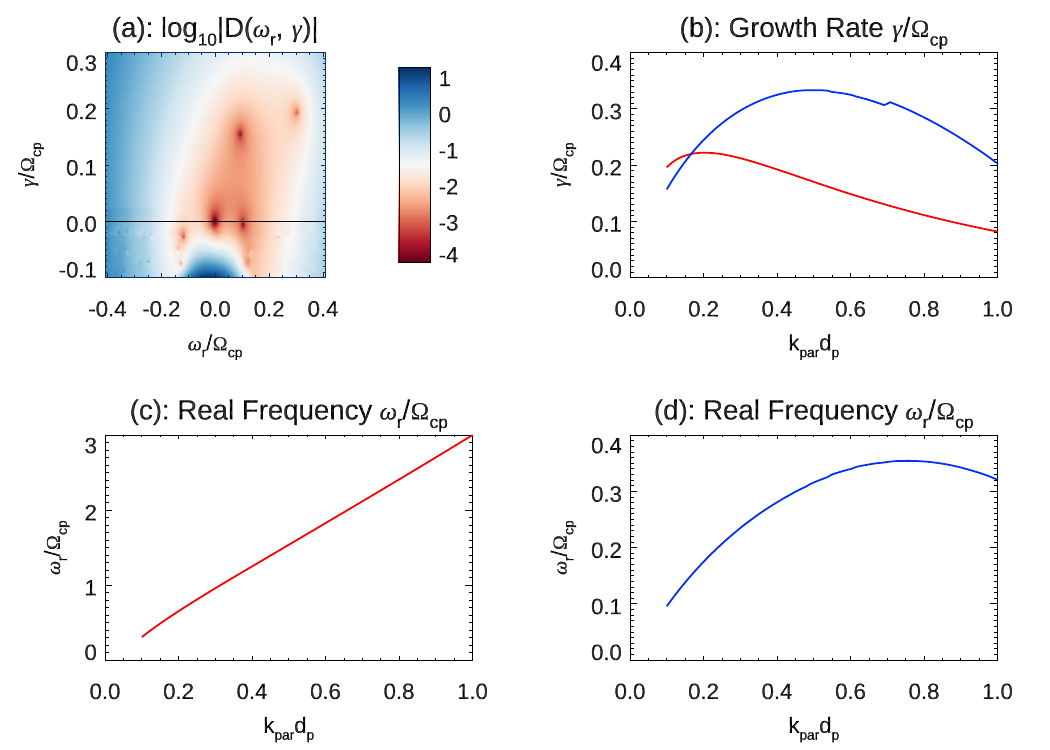}
    \caption{(a) The map of $|D(\omega_r,\gamma)|$ for Case 3 at $k_\parallel = 0.1$, 
    $k_\perp = 10^{-4}$, with minima plotted in dark red. The right-handed mode is 
    located at ($\omega_r/\Omega_{cp}$, $\gamma/\Omega_{cp}$) = (0.31, 0.20) and the left-handed 
    mode is located at ($\omega_r/\Omega_{cp}$, $\gamma/\Omega_{cp}$) = (0.096, 0.16). The (b) 
    growth rates and (c)-(d) real frequencies of the two unstable solutions are plotted as 
    functions of $k_\parallel$, with the right-handed mode plotted in red and the left-handed mode 
    plotted in blue.}
    \label{fig:parmodes_hot}
\end{figure}

Changing the initial energy of the streaming population primarily affects the lower 
frequency, left-handed mode. In the lowest energy case, this mode becomes stable 
everywhere except for a small range around $k_{\parallel}d_p = 0.13$ 
(Fig.~\ref{fig:parmodes_cold}(b)). In the highest energy case, it becomes the dominant mode 
and is strongly unstable for all values of $k_\parallel$, with a peak growth rate of 
$\gamma/\Omega_{cp} = 0.33$ at $k_{\parallel}d_p= 0.48$ (Fig.~\ref{fig:parmodes_hot}(b)). There 
is a modest impact on the right-handed mode as the initial energy is changed, with 
growth rates and real frequencies increasing with beam energy. However, the overall 
characteristics of the mode remain the same across the three cases.

\subsection{Oblique Modes}\label{sec:oblmodes}

\begin{figure}[ht]
    \includegraphics{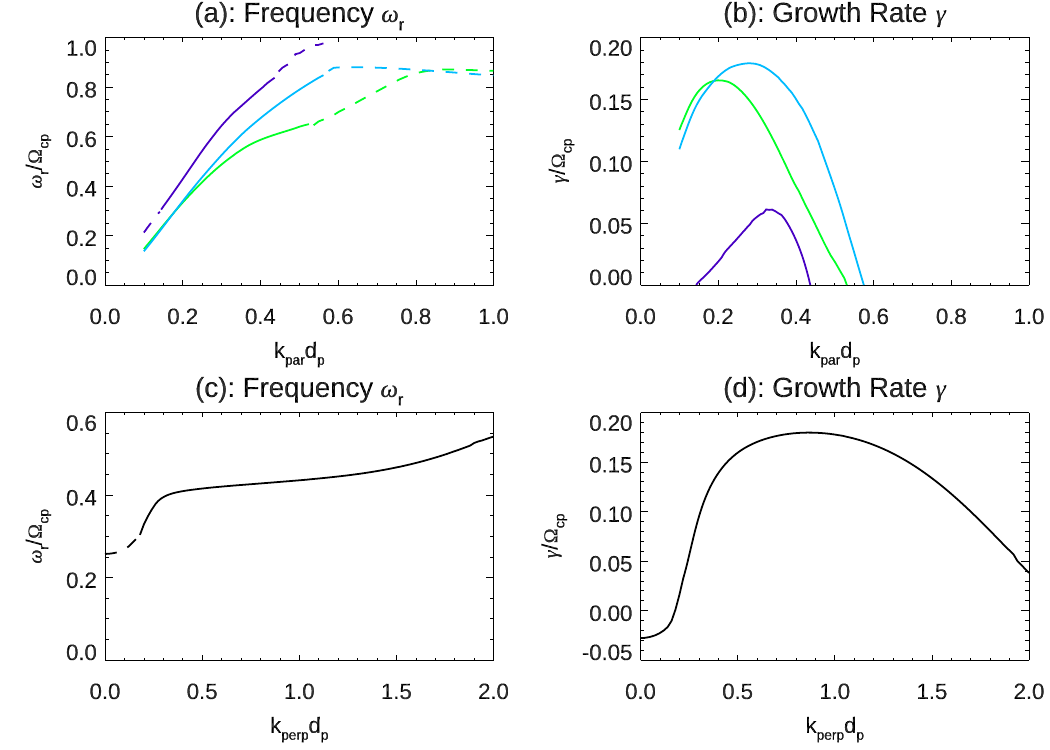}
    \caption{The dispersion properties for the oblique mode in Case 1. The top two plots 
    show, as a function $k_\parallel$, the (a) frequencies and (b) growth rates for 
    $k_{\perp}d_p = 0.5$ (green), 1 (blue), and 2 (purple). Real frequencies while the 
    mode is unstable are plotted as a solid line, while real frequencies while the mode is 
    stable are plotted as a dashed line. The bottom two plots show, 
    as a function of $k_\perp$, the (c) frequency and (d) growth rate for 
    $k_\parallel = 0.25$.}
    \label{fig:oblmodes}
\end{figure}

Using the same method as described for the parallel case, we also find the unstable modes with 
$\Vec{k} \times \Vec{B} \neq 0$. In Case 1, as $k_\perp$ increases from 0, the growth rates for the two 
unstable parallel modes drop off quickly and the lightly damped  mode located at 
($\omega_r/\Omega_{cp}$, $\gamma/\Omega_{cp}$) = (0.1, -0.01) in Fig. \ref{fig:parmodes}(a) 
becomes unstable. It reaches a maximum growth rate of $\gamma/\Omega_{cp}$ = 0.18 around 
$k_{\parallel}d_p$ = 0.28, $k_{\perp}d_p$ = 1, $\omega_r/\Omega_{cp}$ = 0.49 
(Fig. \ref{fig:oblmodes}), becoming stable above $k_{\parallel}d_p$ = 0.6. The 
growth rate is large over a broad range of $k_\perp$ and $\partial\omega_r/\partial k_\perp$
is small around peak growth. At $k_\perp$ = 0, the mode is electrostatic. However, for 
$k_\perp \neq 0$, it gains an electromagnetic component. The electromagnetic field is strong for long 
wavelengths ($|k|d_p < 1$), but weakens as $k_\perp$ increases. At maximum growth, the energy in the 
electric field transverse to $\Vec{k}$ compared to the total wave electric field is $E_T^2/E^2 = 0.28$ 
and the mode is predominately linearly polarized in the perpendicular direction. The increase in the frequency at large $k_\perp$ combined with its electromagnetic (electrostatic) character at long (short) wavelength suggest that this instability can be characterized as a kinetic Alfv\'en wave.

\begin{figure}[ht]
    \includegraphics{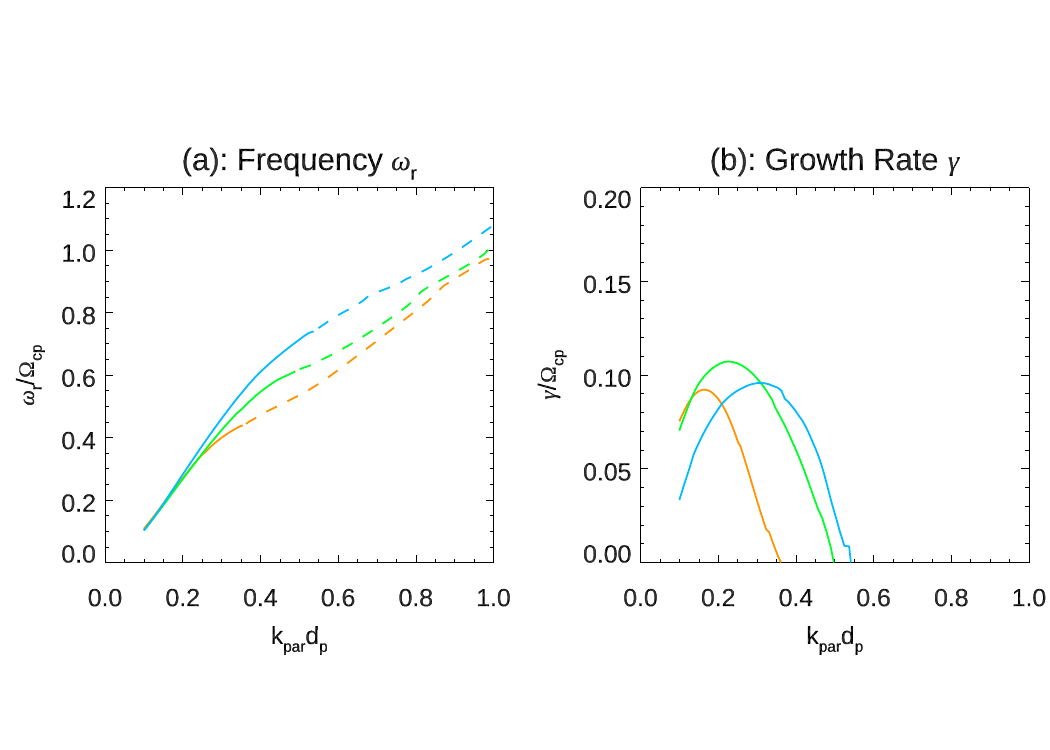}
    \caption{The dispersion properties for the oblique mode in Case 2. The plots show, 
    as a function of $k_\parallel$, the (a) frequencies and (b) growth rates for 
    $k_{\perp}d_p = 0.25$ (orange), 0.5 (green), and 1 (blue). Real frequencies while the 
    mode is unstable are plotted as a solid line, while real frequencies while the mode is
    stable are plotted as a dashed line.}
    \label{fig:oblmodes_cold}
\end{figure}

\begin{figure}[ht]
    \includegraphics{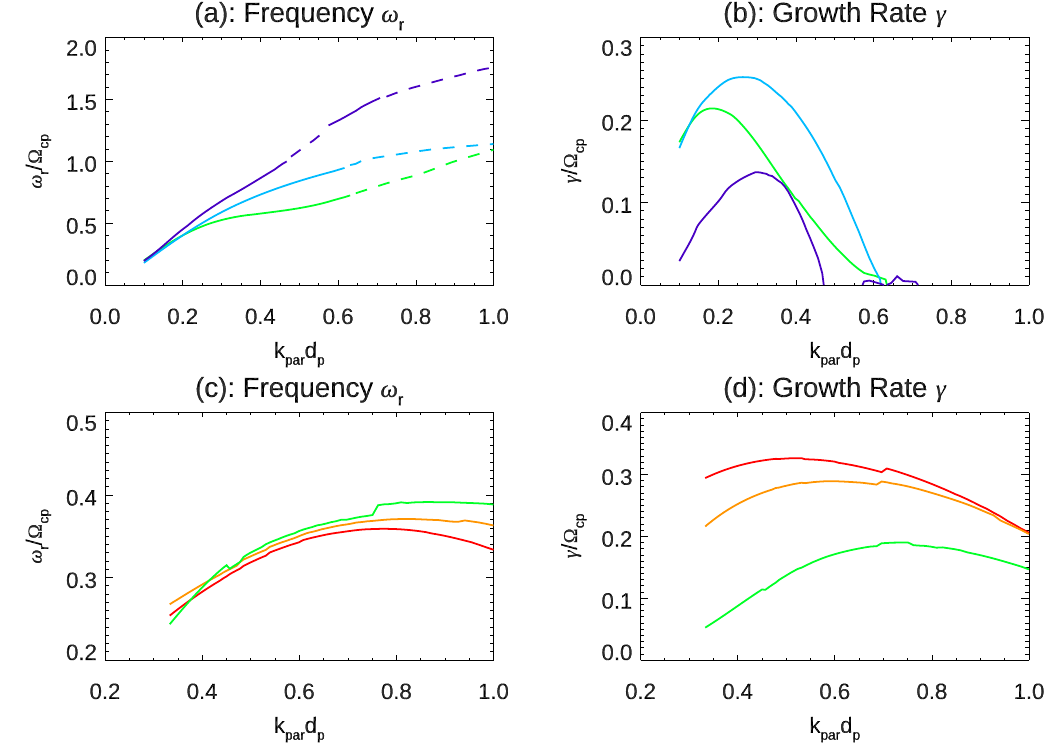}
    \caption{The dispersion properties for both oblique modes in Case 3. The top two plots 
    show, as a function of $k_\parallel$, the (a) frequencies and (b) growth rates for 
    $k_{\perp}d_p = 0.5$ (green), 1 (blue), and 2 (purple) of the linearly polarized 
    mode. Real frequencies while the mode is unstable are plotted as a solid line, while 
    real frequencies while the mode is stable are plotted as a dashed line. The bottom two 
    plots show, as a function of $k_\parallel$, the (c) frequencies and (d) growth rates for 
    $k_{\perp}d_p = 0.1$ (red), 0.25 (orange), and 0.5 (green) of the left-hand polarized 
    mode.}
    \label{fig:oblmodes_hot}
\end{figure}

Lowering the initial energy decreases both the growth rate and obliquity of this mode, 
which, for Case 2, has a maximum growth rate of $\gamma/\Omega_{cp} = 0.11$ 
around $k_{\parallel}d_p = 0.23$, $k_{\perp}d_p = 0.5$, $\omega_r/\Omega_{cp} = 0.31$ 
(Fig. \ref{fig:oblmodes_cold}). Raising the particle energy increases the growth 
rate, reaching a maximum value of $\gamma/\Omega_{cp} = 0.25$ at $k_{\parallel}d_p = 0.26$, 
$k_{\perp}d_p = 1$, $\omega_r/\Omega_{cp} = 0.52$ in Case 3 
(Fig. \ref{fig:oblmodes_hot}(a)-(b)). In both cases, the value of $E_T^2/E^2$ at peak growth is 
the same as in Case 1.

In the highest energy case, there are also significant growth rates for the left-handed 
mode discussed in Sec. \ref{sec:parmodes} for $k_\perp \neq 0$ 
(Fig. \ref{fig:oblmodes_hot}(c)-(d)). Although the parallel wavenumber associated with the peak 
growth rate increases as $k_\perp$ increases, the other characteristics of the mode 
remain the same as in the $k_\perp = 0$ case.

\subsection{Impact of $^4$He}

\begin{figure}[ht]
    \includegraphics{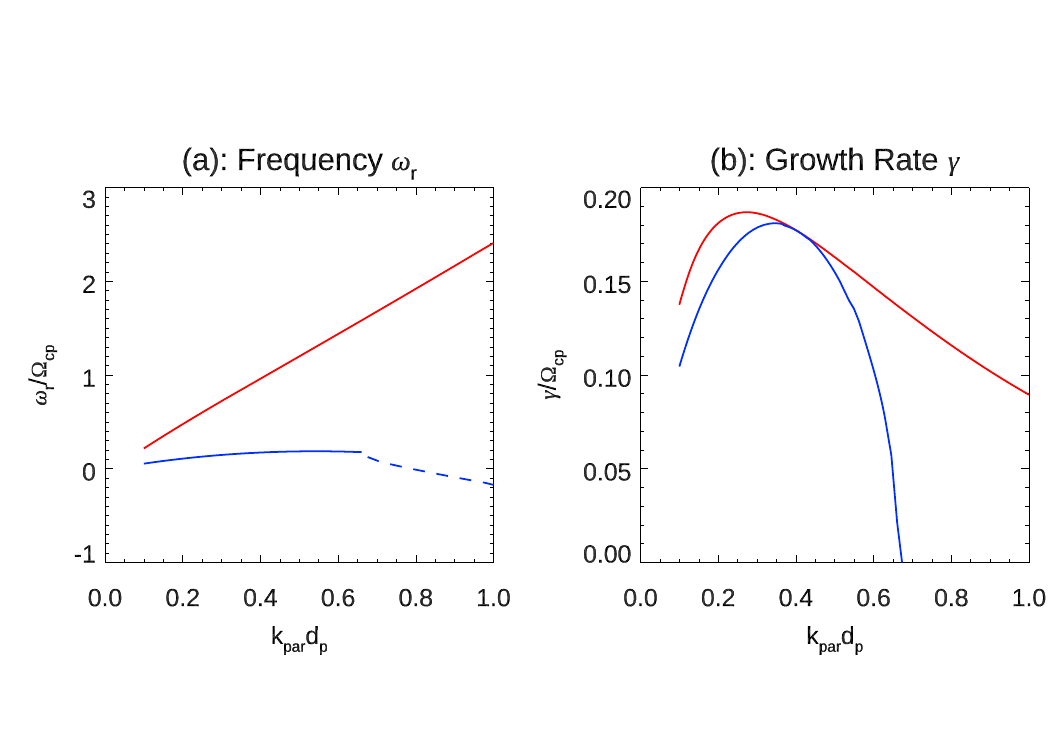}
    \caption{The (a) frequencies and (b) growth rates as a function of $k_\parallel$ for the 
    parallel modes in Case 1 including a $^4$He number density equal to 5\% that of the 
    protons (compare to Fig.~\ref{fig:parmodes} with no $^4$He). The right-hand polarized 
    mode is plotted in red and the left-handed mode is 
    plotted in blue. Real frequencies while the mode is unstable are plotted as a solid 
    line, while real frequencies while the mode is stable are plotted as a dashed line.}
    \label{fig:parmodes_alph}
\end{figure}

\begin{figure}[ht]
    \includegraphics{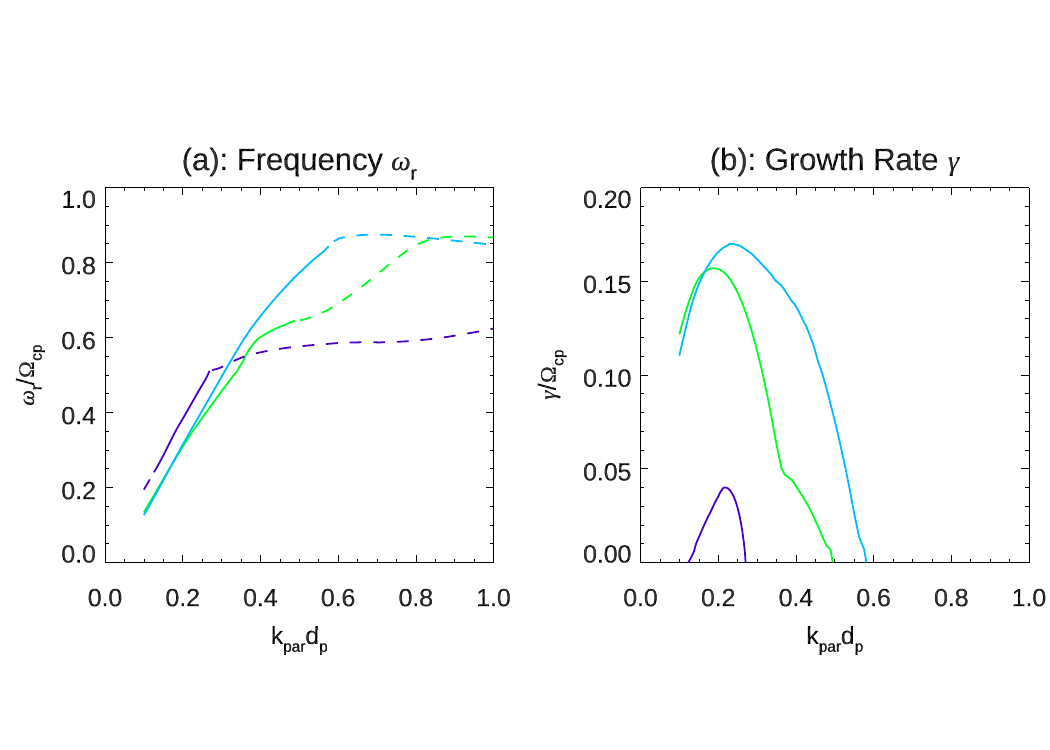}
    \caption{The (a) frequencies and (b) growth rates as a function of $k_\parallel$ for 
    $k_\perp$ = 0.5 (green), 1 (blue), and 2 (purple) in Case 1 including a $^4$He number 
    density equal to 5\% that of the protons (compare to Fig.~\ref{fig:oblmodes} with no $^4$He). 
    Real frequencies while the mode is unstable 
    are plotted as a solid line, while real frequencies while the mode is stable are plotted 
    as a dashed line.}
    \label{fig:oblmodes_alph}
\end{figure}

In order to test the impact of $^4$He on the spectrum of modes, we repeat the linear analysis 
while including a population of $^4$He with a number density equal to 5\% that of the protons, 
reflecting typical abundances in the corona. The initial distribution functions 
are given by Eq. \ref{eq:distfunc}, with the same temperatures in each case as the 
protons. As a representative example, the frequencies and growth rates of the unstable 
modes for Case 1 are shown in Figure \ref{fig:parmodes_alph} (parallel modes) and Figure 
\ref{fig:oblmodes_alph} (oblique modes). Since there is very little difference between 
these results and those presented in the previous sections, we conclude that, at the low 
densities expected in the corona, this species has little impact on the generation of ion beam-driven 
instabilities.

\section{Discussion}\label{sec:disc}

The linear theory results from {\tt ALPS} are in good agreement with the simulation results 
presented in FDS.\cite{fitzmaurice_wave_2024} The linear parallel right-handed mode that is 
unstable in all cases is the primary cause for the parallel waves observed in the 
simulations for Cases 1 and 2 and at late time for Case 3. It is most likely a form of 
the right-hand resonant instability that is well documented for ion beams 
(see Sec. \ref{sec:intro}). 

The large growth rates for the left-handed mode in Case 3 lead to the higher parallel 
wavenumbers and mixed polarity observed at early time in the simulations for this case. 
It is less certain if it can be identified with any of the instabilities previously discussed in 
the literature. Given that the growth rates are strongly dependent on the beam 
energy and that the mode remains unstable for oblique wavevectors, it appears to be similar 
to the Alfv\'en instabilities discussed in \citet{daughton_electromagnetic_1998}.

In the PIC simulation results of FDS,\cite{fitzmaurice_wave_2024} the waves produced by the distributions in Case 2 are primarily 
parallel, which is consistent with the smaller linear growth rates for the oblique mode in 
this case. For Cases 1 and 3, linear theory predicts large growth rates for this mode 
at nearly perpendicular wavenumbers ($\approx 75^{\circ}$). This is reproduced well in the simulation 
results for both cases at early time. The mode is most likely a kinetic Alfv\'en wave, since the 
electromagnetic field is strong in the long wavelength limit but weakens as 
$k_\perp d_p \sim k_\perp \rho_s$ approaches 1. 

According to the linear theory, interactions between the waves and the ions will occur when the 
velocity of a particle is equal to the resonant velocity, $v_r = (\omega_r \pm n\Omega_{cp})/k_\parallel$, 
where the plus sign corresponds to electrostatic and right-hand polarized waves and the minus sign to 
left-handed waves. In the case of parallel propagation for electrostatic waves, $n = 0$ and the resonant 
velocity equals the phase velocity of the wave, so that wave-particle interactions only occur through 
the Landau resonance. For parallel propagating electromagnetic waves, $n = 1$; therefore, interactions 
occur at the cyclotron resonance, which is above (below) the phase velocity for right-handed 
(left-handed) waves. For all wave modes at oblique propagation, there are an infinite number of resonant 
velocities ($n = 0$, $\pm 1$, $\pm 2$, ...), both above and below the phase velocity of the wave.

From the linear resonance condition, all three instabilities will be important for proton and alpha scattering 
during solar energy release. The parallel right-handed mode and the positive resonances of the kinetic 
Alfv\'en wave will scatter particles in the high energy tail of the distribution, reducing their parallel 
energy and converting it into perpendicular energy. The parallel left-handed mode and the negative 
resonances of the kinetic Alfv\'en wave will scatter particles into negative parallel velocities, 
contributing to thermalization of the distribution. The strong scattering of protons and alpha particles resulting from these instabilities will cause these particles to diffuse rather than free-stream along the magnetic field within the flaring region. This will greatly increase the time required for them to escape from the flare energy release region and  therefore extend the time over which particles can gain energy, which will facilitate greater energy gain.

Heating cold $^3$He requires a resonance at $v_\parallel = 0$, which is not possible for the parallel 
right-handed mode due to the positive definite numerator in $v_r$ when $\omega_r > 0$. For the parallel 
left-handed mode, this would require $\omega_r/\Omega_{cp} = 2/3$, which is much higher than the frequencies 
predicted for this mode by the linear theory (see Figs.~\ref{fig:parmodes}-\ref{fig:parmodes_hot}). The kinetic Alfv\'en wave will contribute to $^3$He 
heating through the $n = -1$ resonance, since it has unstable frequencies near the resonance frequency of $^3$He: 
$\omega_r/\Omega_{cp} = 2/3$ (see Figs.~\ref{fig:oblmodes}-\ref{fig:oblmodes_hot}).

However, wave heating of $^3$He is not limited to those waves with $\omega_r/\Omega_{cp} = 2/3$. As discussed in \citet{temerin_production_1992} and FDS,\cite{fitzmaurice_wave_2024} the 
frequencies of waves traveling through the corona will remain constant while the cyclotron 
frequencies of the constituent ion species will change with the magnetic field strength, 
according to $\Omega_{ci} = qB/mc$. Therefore, waves with frequencies 
$1/2 < \omega_r/\Omega_{cp} < 2/3$ will heat $^3$He above the flare site where $|B|$ is smaller. For $\omega_r/\Omega_{cp}<1/2$, other ion species will dominate absorption compared with $^3$He because of the latter's small number density.  Waves with 
frequencies $2/3 < \omega_r/\Omega_{cp} < 1$ will heat $^3$He below the flare site where $|B|$ is larger. Frequencies above this range will be absorbed by protons. Because of their increased thermal speeds, heated $^3$He ions 
can then migrate back into the energy release site, increasing their local abundance. Within the flaring region this $^3$He will be further 
accelerated to high energy by the usual Fermi reconnection drive mechanism \citep{yin_simultaneous_2024}. From the linear theory results for 
Cases 1 and 3, the frequencies at peak growth are in the range $1/2< \omega_r/\Omega_{cp} < 1$, so there should 
be substantial energy in this frequency range to heat $^3$He. Since the addition of $^4$He had little 
effect on the linear analysis, we conclude that kinetic Alfv\'en waves driven by reconnection-accelerated protons 
streaming out from energy release sites are a strong candidate for driving the $^3$He enhancements seen in impulsive events.

\section{Acknowledgments}
The authors were supported by NSF grant PHY2109083, NASA grants 80NSSC20K1813 and 80NSSC20K1277, and NASA FINESST award 80NSSC23K1625. This study benefits
from discussions within the International Space Science
Institute (ISSI) Team ID 425 “Origins of 3He-rich SEPs.” We acknowledge informative discussions with Drs. R.\ Bucik and G.\ Mason. {\tt ALPS} can be found at \url{github.com/danielver02/ALPS} and assistance in using the code was provided by Dr. K.\ Klein. 

\section{Author Declarations}
\subsection{Conflicts of Interest}
The authors have no conflicts to disclose.

\subsection{Author Contributions}
\textbf{A. Fitzmaurice}: Conceptualization (supporting); formal analysis (lead); funding acquisition 
(equal); investigation (lead); methodology (equal); visualization (lead); writing--original draft (lead); 
writing--review and editing (equal). \textbf{J. F. Drake}: Conceptualization (lead); formal analysis 
(supporting); funding acquisition (equal); methodology (equal); project administration (lead); supervision 
(lead); writing--original draft (supporting); writing--review and editing (equal). \textbf{M. Swisdak}: 
Conceptualization (supporting); formal analysis (supporting); funding acquisition (equal); methodology 
(equal); supervision (supporting); writing--original draft (supporting); writing--review and editing 
(equal).

\section{Data Availability}
The data that support the findings of this study are available from the corresponding author upon reasonable request.

\bibliography{mylibrary}

\end{document}